\documentclass{article}
\usepackage{spconf,amsmath,graphicx,hyperref}
\usepackage{amssymb}
\usepackage{cite}
\usepackage{xcolor}
\usepackage{pifont}
\usepackage{booktabs}
\usepackage{multirow}
\usepackage{cleveref}
\usepackage{fancyhdr}

\title{PicoAudio2: Temporal Controllable Text-to-Audio Generation with Natural Language Description}
\name{Zihao Zheng$^{\dag 1,2}$\thanks{$\dag$Work done partially during internship at Shanghai AI Lab}, Zeyu Xie$^{1}$, Xuenan Xu$^{1,2}$, Wen Wu$^{2}$, Chao Zhang$^{2}$, Mengyue Wu$^{*1}$\thanks{*Mengyue Wu is the corresponding author.}}
\address{
    $^{1}$MoE Key Lab of Artificial Intelligence, X-LANCE Lab, Shanghai Jiao Tong University \\
    $^{2}$Shanghai AI Lab
}
\begin{document}
\ninept
\maketitle
\begin{abstract}
While recent work in controllable text-to-audio (TTA) generation has achieved fine-grained control through timestamp conditioning, its scope remains limited by audio quality and input format.
These models often suffer from poor audio quality in real datasets due to sole reliance on synthetic data.
Moreover, some models are constrained to a closed vocabulary of sound events, preventing them from controlling audio generation for open-ended, free-text queries. 
%
This paper introduces PicoAudio2, a framework that advances temporal-controllable TTA by mitigating these data and architectural limitations.
Specifically, we use a grounding model to annotate event timestamps of real audio-text datasets to curate temporally-strong real data, in addition to simulation data from existing works.
The model is trained on the combination of real and simulation data.
Moreover, we propose an enhanced architecture that integrates the fine-grained information from a timestamp matrix with coarse-grained free-text input.
Experiments show that PicoAudio2 exhibits superior performance in terms of temporal controllability and audio quality.
The demo page can be accessed at \url{https://HiRookie9.github.io/PicoAudio2-Page}.
\end{abstract}
\section{Introduction}
\label{sec:intro}

Recently, text-to-audio generation (TTA) has garnered significant research interest due to its broad application in fields such as virtual reality and social media content creation.
With the emergence of advanced generative models~\cite{ho2020denoising,lipman2023flow}, TTA has made progress in generating high-fidelity audio from textual descriptions~\cite{liu2024audioldm2,ghosal2023tango,majumder2024tango,huang2023maa2,hai2024ezaudio}.
Despite these improvements, fine-grained temporal control remains a critical and understudied challenge.
Unlike text-to-speech tasks, where alignment is naturally structured, general audio datasets lack fine-grained temporal annotations~\cite{kim2019audiocaps, drossos2020clotho, Mei_2024}, and text encoders often fail to capture precise temporal cues.
As a result, mainstream TTA models struggle to generate audio recordings that are precisely aligned with the input text.
Several recent works have begun exploring temporal controllability in TTA.
For example, Make-An-Audio 2 (MAA2)~\cite{huang2023maa2} leverages large language models (LLMs) to convert free-text input into structured temporal descriptions (e.g., an event happens throughout, at the beginning, or at the end).
However, such control remains coarse-grained.
AudioComposer~\cite{wang2024audiocomposer} also employs structured text to encode control information such as temporal occurrence and energy and achieves high control accuracy. However, captions in the training data are all simulated, resulting in limited audio quality on real-world datasets like AudioCaps~\cite{kim2019audiocaps}, which will be shown in \Cref{subsec:generation_performance}.
PicoAudio~\cite{xie2025picoaudio} introduces a timestamp matrix to indicate the occurrence time of events, achieving high audio quality and temporal controllability of pre-defined short-duration events.
However, as the timestamp matrix is limited to pre-defined categories, it lacks the flexibility to accommodate natural language descriptions and open-ended event specifications.


\begin{table}[t]
    \centering
    \scriptsize
    \begin{tabular}{l|ccc}
        \toprule
        \textbf{Model} & Free Text & Audio Quality & Fine Temporal Control \\
        \midrule
        MAA2 et al.~\cite{liu2024audioldm2,majumder2024tango,huang2023maa2}      & \textcolor{green}{\ding{51}}    & \textcolor{green}{\ding{51}} & \textcolor{red}{\ding{55}} \\
        AudioComposer~\cite{wang2024audiocomposer}       & \textcolor{green}{\ding{51}}  & \textcolor{red}{\ding{55}} & \textcolor{green}{\ding{51}} \\
        PicoAudio ~\cite{xie2025picoaudio}            & \textcolor{red}{\ding{55}}   & \textcolor{green}{\ding{51}} & \textcolor{green}{\ding{51}} \\
        PicoAudio2 (ours)     & \textcolor{green}{\ding{51}}  & \textcolor{green}{\ding{51}} & \textcolor{green}{\ding{51}} \\
        \bottomrule
    \end{tabular}
    \caption{Comparison of models with respect to free text, audio quality, and temporal control.}
    \label{tab:model_comparison}
\end{table}

\begin{figure*}[ht]
    \centering
    \includegraphics[width=0.95\textwidth]{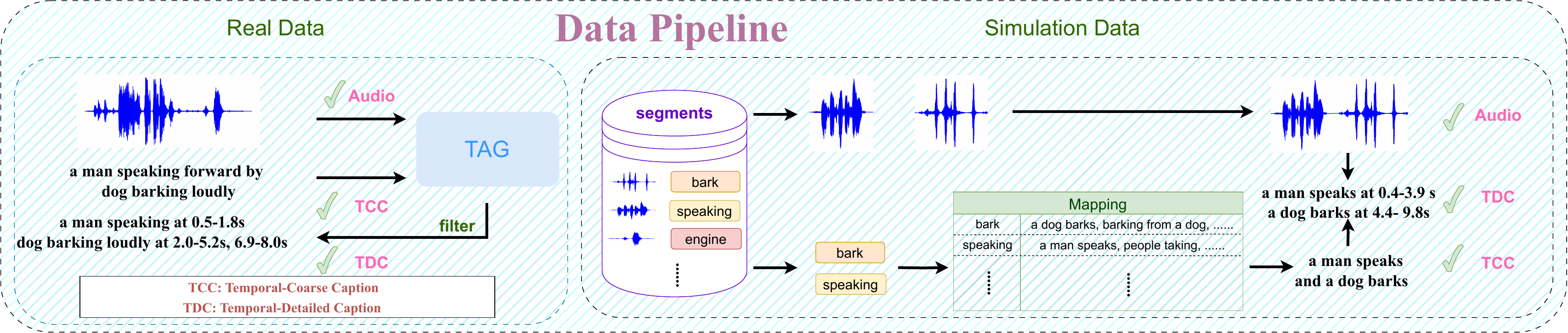}
    \caption{The data curation pipeline. The left part shows the real dataset processing pipeline, where the TAG model extracts event timestamps and data with omissions or overlaps are excluded. The right part shows the data simulation pipeline, where multi-event audio is simulated from preprocessed single-event segments with precise timestamp information. Captions are obtained by concatenating single-event descriptions.}
    \label{fig:data_pipeline}
\end{figure*}

To achieve precise temporal controllability and high audio quality given free-text input, we propose PicoAudio2, as shown in \Cref{tab:model_comparison}. 
PicoAudio2 incorporates improvements in both data and model design, supporting temporal control with free-text input. 
In terms of data, we develop dedicated data processing pipelines for both simulation and real data.
For simulation data, we synthesize audio-caption-timestamp triplets following the AudioTime~\cite{xie2025audiotime} method.
Since the original event labels are categorical rather than natural language descriptions, we convert these labels into textual descriptions via LLMs and extract embeddings with the Contrastive Language-Audio Pre-training (CLAP) model~\cite{laionclap2023}. 
However, there is still a distribution gap between simulated datasets and real datasets, both in the audio and caption domains.
Therefore, we incorporate real datasets into training to mitigate the gap.
For real datasets, we segment captions into single-event descriptions using LLMs, and extract timestamp information with a Text-to-Audio Grounding (TAG) model \cite{xu2024towards}. 

In terms of model design, PicoAudio2 employs free-text descriptions for individual events to construct a timestamp matrix. 
Single-event descriptions are first transformed to embeddings, which are then combined with corresponding timestamps to form the matrix. This matrix provides fine-grained temporal control signals, while the original coarse-grained caption offers global semantics. 
The integration of both inputs allows the model to achieve precise temporal alignment from open-ended language descriptions. 
In addition, such a design retains training flexibility and allows for the inclusion of data without timestamp annotations by replacing the matrix with a fixed embedding. 

Overall, our contributions are summarized as the following:
\begin{itemize}
    \item We propose a novel data processing pipeline that enhances simulated dataset and incorporates real data into training.
    \item We design a temporal-controllable generation framework, enabling high-quality audio generation with fine-grained timestamp control based on free-text descriptions.
    \item Extensive experiments demonstrate that our framework achieves precise temporal control and high audio quality.
\end{itemize}

\section{Temporally-Aligned Data Curation}
\label{sec:data_curation}
To leverage a larger amount of data for training, we design curation pipelines for both simulation and real data, as shown in \Cref{fig:data_pipeline}.
Typical real datasets such as AudioCaps~\cite{kim2019audiocaps}, consist of audio-caption pairs without precise temporal annotations.
We refer to such captions as temporal coarse captions (TCC), and the corresponding audio–caption pairs as temporally weak data.
Our pipelines are designed to convert them into temporally detailed captions (TDC), containing exact timestamps for each sound event, resulting in audio–TCC–TDC triplets that constitute temporally strong data for training.


\subsection{Simulation Data}
\label{subsec:simulation_data}
We adopt the approach in AudioTime~\cite{xie2025audiotime} to synthesize high-quality audio-TCC-timestamp triplets.
However, the single-event descriptions in TCC are categorical labels (e.g., `barking', `speaking').
Given the gap between category labels and free-form text, we leverage LLMs~\cite{achiam2023gpt} to generate multiple free-text descriptions for each category.
Additionally, we use LLMs to extract single-event descriptions from AudioCaps caption corpus.
Then we assign category labels to these descriptions based on CLAP~\cite{laionclap2023} similarities.
For each category, we first select the top 30 free-text candidates based on CLAP scores, followed by manual review to ensure quality.

Ultimately, each category label is mapped to 15–30 textual descriptions. 
For instance, ``dog'' is mapped to descriptions such as ``a dog is barking and yipping'' and ``barking from a dog''.
During training, categorical labels in TCC are mapped to randomly sampled descriptions from their corresponding candidate lists.
These descriptions are then combined with timestamps to form TDC.
Finally, we construct the audio-TCC-TDC triplets, which serve as temporally strong data.

\subsection{Real Data}
\label{subsec:real_data_pipeline}
For real data where audio and TCC are already available, we first use LLMs to extract single-event descriptions from TCC.
These descriptions are then timestamped with the TAG model~\cite{xu2024towards}, associating single-event descriptions with temporal occurrence to form TDC.
However, we observe that the TAG model exhibits certain limitations: 1) inaccuracy when multiple events overlap; 2) occasionally fails to detect events mentioned in the TCC.
Therefore, to ensure timestamp accuracy, we filter out samples with timestamp overlaps or omissions, retaining only well-curated data as temporally strong data.
In addition, the original audio-TCC pairs are retained as temporally weak data and used during training to enhance generalization.

\section{PicoAudio2}

\begin{figure*}[ht]
    \centering
    \includegraphics[width=0.95\textwidth]{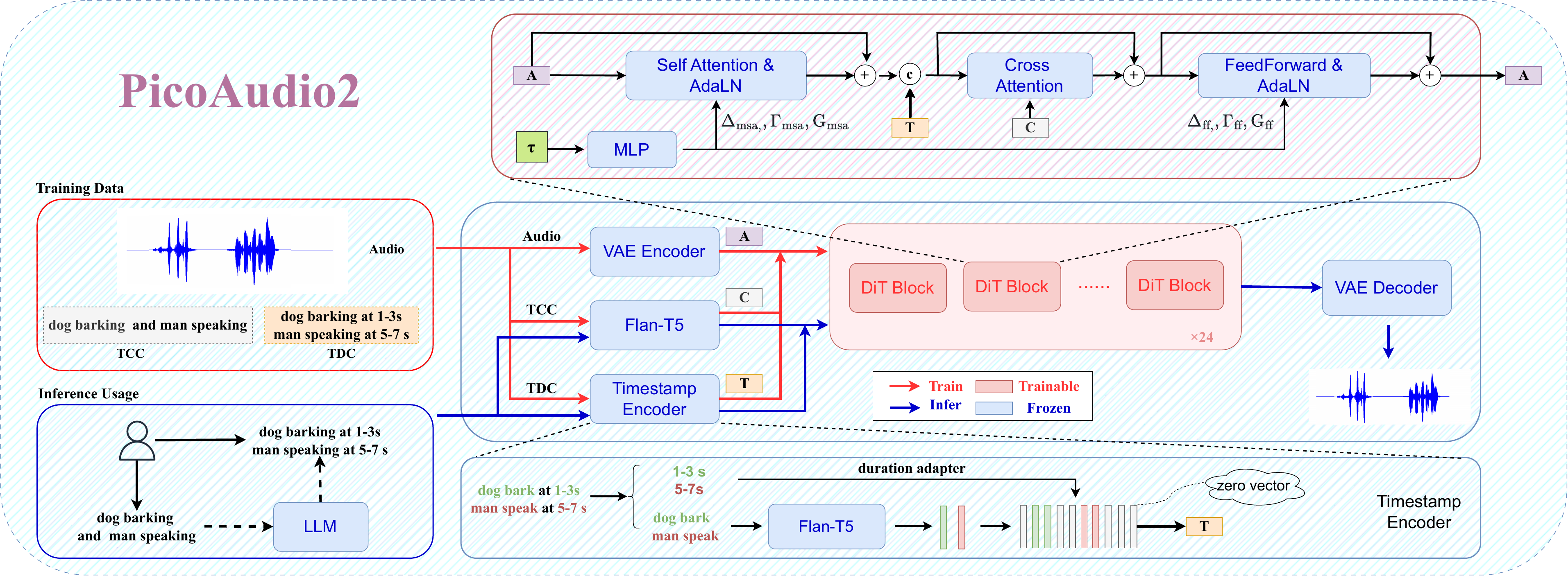}
    \caption{PicoAudio2 framework. The red arrow represents the training process while the blue represents inference. During inference, users can either provide detailed timestamps for each events, or a coarse description for LLM to infer the timestamp information.}
    \label{fig:model}
\end{figure*}

The overall structure of PicoAudio2 is shown in \Cref{fig:model}, which comprises a Variational Autoencoder (VAE), a text encoder, and a timestamp encoder.
The VAE operates on raw waveforms to compress audio signals into latents $\mathbf{A}$ during training and transforms latents back to audio waveforms during inference.
We employ the VAE in EzAudio~\cite{hai2024ezaudio}.
The timestamp encoder extracts temporal information from TDC, obtaining timestamp matrix $\mathbf{T}$.
The Flan-T5 \cite{chung2024scaling} text encoder extracts semantic information $\mathbf{C}$ from TCC.
VAE and Flan-T5 are frozen during training.


\subsection{Timestamp Encoder}
The timestamp encoder extracts the timestamp matrix $\mathbf{T}$ from TDC to provide temporal information.
For temporally weak data (which lack TDC), $\mathbf{T}$ is set as a fixed embedding sequence.
This design enables PicoAudio2 to generate valid audio with only TCC input.
For temporally strong data, the timestamp encoder first transforms single-event descriptions in TDC to event-level features $\mathbf{a} \in \mathbb{R}^C$ using Flan-T5.
Then these features are aggregated according to timestamps of each event, resulting in $\mathbf{T} \in \mathbb{R}^{T\times C}$: 
\begin{align*}
 \mathbf{T}_t=\begin{cases}
 \sum_{i}  \mathbf{a}_i & \text{ if event i occurs at t}  \\
 \mathbf{0} & \text{otherwise} 
\end{cases}
\end{align*}
For each timestep $t$, features of occurring events are summed as $\mathbf{T}_t$.
In contrast to PicoAudio~\cite{xie2025picoaudio}, which maps textual descriptions to fixed classes, PicoAudio2 is capable of handling any events described by natural language while still providing the representative timestamp matrix $\mathbf{T}$ that remains temporally aligned with the audio latent $\mathbf{A}$.

During training, TDC are obtained either through simulation or by preprocessing real data.
During inference, users can provide TDC directly, or provide TCC and rely on LLMs to transform TCC to TDC.
Since LLMs are sufficiently powerful to generate reasonable TDC from TCC, we focus on guiding the model to generate audio recordings that adhere to the temporal constraints specified in TDC.

\subsection{Diffusion Backbone}

The backbone is a Diffusion Transformer (DiT)~\cite{hai2024ezaudio} consisting of several DiT blocks.
Similar to a standard Transformer block, each DiT block consists of a self-attention layer, a cross-attention layer and a feedforward network (FFN) layer.
Residual connection is applied after self-attention, cross-attention and FFN.
AdaLN is used to fuse diffusion timestep $\mathbf{\tau}$ before self-attention and FFN.

The audio latent $\mathbf{A}$ is first fed to the self-attention layer:
\begin{align*}
    \mathbf{A} &= \mathrm{AdaLN_1}(\mathbf{A}, \mathbf{\tau})\\
    \mathbf{A} &= \mathrm{Attn}(\mathbf{A}, \mathbf{A}, \mathbf{A}) + \mathbf{A}
\end{align*}

Between self-attention and cross-attention, $\mathbf{T}$ is fused by concatenation with $\mathbf{A}$ since they are aligned along the time axis.

\begin{align*}
    \mathbf{A} = \mathrm{Concat}(\mathbf{A}, \mathbf{T})
\end{align*}

Then, $\mathbf{C}$ is integrated via cross-attention, followed by FFN:
\begin{align*}
     \mathbf{A} &= \mathrm{Attn}(\mathbf{A}, \mathbf{C}, \mathbf{C}) + \mathbf{A} \\
     \mathbf{A} &= \mathrm{AdaLN_2}(\mathbf{A}, \mathbf{\tau})\\
     \mathbf{A} &= \mathrm{FFN}(\mathbf{A}) + \mathbf{A}
\end{align*}
The model is trained by standard diffusion loss with the velocity target~\cite{salimansprogressive}.
During inference, classifier-free guidance~\cite{ho2021classifier} is used.

\section{Experiments}
\label{sec:pagestyle}

\begin{table*}[t]
 \centering
 \caption{General audio quality and temporal controllability results. AudioCaps-DJ (DisJoint) is the subset of AudioCaps without timestamp omissions or overlaps. Best results are in bold.}
 \label{tab:main_result}
 \small 
 \begin{tabular}{ll|cccc|c|cc|c}
 \toprule
  & & \multicolumn{5}{c}{\textbf{General Audio Quality}} & \multicolumn{3}{|c}{\textbf{Temporal Accuracy}} \\
 \cmidrule{3-10}
 \multicolumn{2}{c|}{} & \textbf{FD}$\downarrow$ & \textbf{KL}$\downarrow$ & \textbf{IS}$\uparrow$ & \textbf{CLAP}$\uparrow$ & 
 \textbf{MOS-Q}$\uparrow$ & \textbf{Seg-F$_1$}$\uparrow$  & 
 \textbf{Seg-F$_1$-ME}$\uparrow$& 
 \textbf{MOS-T}$\uparrow$ \\
 \midrule
 \multirow{7}{*}{AudioCaps-DJ} & AudioLDM2~\cite{liu2024audioldm2}& \textbf{28.982} & 2.447 & 9.333 & 0.340& 2.77& 0.644& 0.396& 2.05\\
 & Tango2~\cite{majumder2024tango} & 37.315&2.534 & 10.844 & 0.365 & \textbf{3.49}&0.659& 0.433& 2.90\\
 & MAA2~\cite{huang2023maa2} & 43.407 & \textbf{2.364} & 9.427 & 0.351& 3.30& 0.647& 0.434& 2.60\\
 & AudioComposer~\cite{wang2024audiocomposer} & 46.833 &3.002 & 6.202 & 0.254& 2.47&0.690& 0.613 & 3.80\\
 \cmidrule{2-10}
 & PicoAudio2 (w/o T) &37.861	&2.626	&11.610	&\textbf{0.373}& 2.83	&0.659 & 0.432& 2.42\\
 & PicoAudio2 & 39.961&	2.618	&\textbf{12.253}&	0.370& 3.29&\textbf{0.857} & \textbf{0.771}& \textbf{4.15}\\
 \bottomrule
 \end{tabular}
\end{table*}
\begin{table}[ht]
 \centering
 \caption{General audio quality results on the test set of AudioCaps. Best results are in bold.}
 \label{tab:audiocaps_result}
 \small
 \begin{tabular}{l|cccc}
 \toprule
  & \textbf{FD}$\downarrow$ & \textbf{KL}$\downarrow$ & \textbf{IS}$\uparrow$ & \textbf{CLAP}$\uparrow$ \\
 \midrule
 AudioLDM2~\cite{liu2024audioldm2} & \textbf{22.598} & \textbf{2.252} & 9.914 & 0.334 \\
 Tango2~\cite{majumder2024tango} & 36.301 & 2.304 & 9.886 & \textbf{0.392} \\
 MAA2~\cite{huang2023maa2} & 26.699 & 2.335 & 10.392 & 0.372 \\
 AudioComposer~\cite{wang2024audiocomposer} & 27.418 & 2.516 & 7.864 & 0.285 \\
 PicoAudio2 & 27.388 & 2.451 & \textbf{12.347} & 0.383 \\
 \bottomrule
 \end{tabular}
\end{table}

\paragraph*{Datasets}
Both simulation and real data are used for training.
For simulation data, we generate 64K audio clips. 
Each clip lasts at most 10 seconds and includes one to four events.
For real data, we transform captions in AudioCaps~\cite{kim2019audiocaps} and WavCaps-ASSL subset~\cite{Mei_2024} to TDC, resulting in 49K high-quality audio-TDC pairs. We also utilize around 106K temporally weak data samples from these two datasets.
Overall, the total training data consists of two parts: \textasciitilde106K temporally weak data (in the form of audio-TCC pairs) and \textasciitilde113K temporally strong data (in the form of audio-TCC-TDC triplets).
During training, a sampling ratio of 1:2 between temporally weak and strong data is used.

\paragraph*{Experimental Setup}
The time resolution of the timestamp matrix is set to 20 ms.
The diffusion backbone follows EzAudio~\cite{hai2024ezaudio}, with 24 layers, 16 attention heads and a hidden size of 1024.
PicoAudio2 is trained for 50 epochs with a maximum learning rate of $1 \times 10^{-4}$.
The learning rate takes a linear decay schedule with a weight decay of $1 \times 10^{-6}$. 
During inference, the classifier-free guidance scale is set to 7.5.

\paragraph*{Evaluation}

The evaluation is performed in terms of general audio quality and temporal controllability.
For general audio quality, Frechet Distance (FD), Inception Score (IS), Kullback–Leibler divergence (KL) and CLAP score are used.
For temporal controllability, we annotate the timestamps of each event using TAG~\cite{xu2024towards}.
As described in \Cref{subsec:real_data_pipeline}, we retain data without timestamp omissions or overlaps, naming this subset AudioCaps-DisJoint (DJ).
We use Segment-F$1$ (Seg-F$_1$)~\cite{2016Metrics} as the metric.
Since the precise temporal control of multiple events is more difficult than single events, we filter AudioCaps-DJ to construct a multi-event subset (exclusively for temporal controllability evaluation), reporting the corresponding Seg-F$1$-ME (Multi-Event).

Besides objective metrics, we report the Mean Opinion Score (MOS): MOS-Q for audio quality and MOS-T for temporal controllability. 10 high-educated raters without hearing loss are invited to score 10 random samples from AudioCaps-DJ.

\section{Results}

\begin{figure}[ht]
    \centering
    \includegraphics[width=0.45\textwidth]{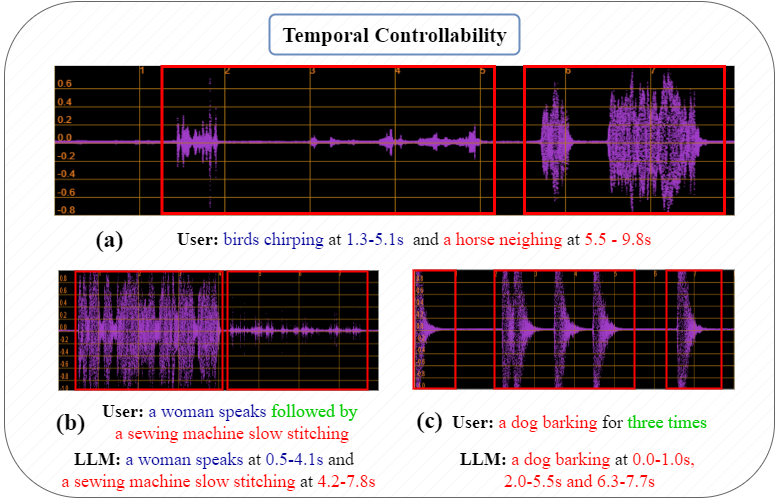}
    \caption{During inference, users can provide TDC like (a), or TCC, which will be transformed to TDC like (b) and (c).}
    \label{fig:demo}
\end{figure}

\subsection{Generation Performance}
\label{subsec:generation_performance}

\paragraph*{General Audio Quality}
We first compare PicoAudio2 with baselines, including AudioLDM2~\cite{liu2024audioldm2}, Tango2~\cite{majumder2024tango}, MAA2~\cite{huang2023maa2} and AudioComposer~\cite{wang2024audiocomposer}, on general audio quality.
Among them, MAA2 and AudioComposer require specified input formats to achieve temporal control so the inputs for these two models are constructed accordingly.
For Tango2 and AudioLDM2, we input TCC and TDC respectively, and take the better-performing result for evaluation. 
Results are shown in \Cref{tab:main_result} and \ref{tab:audiocaps_result}.
Per subjective and objective metrics, PicoAudio2 achieves performance comparable to leading models on both AudioCaps and AudioCaps-DJ, indicating high audio quality. 
On AudioCaps-DJ, AudioComposer performs notably worse.
It takes both event and temporal information within its input descriptions, which may lead to suboptimal event fidelity.
On AudioCaps, AudioComposer achieves better audio quality without timestamp input, further validating the negative influence of temporal information.
By decoupling captions with temporal information, PicoAudio2 is capable of generating high-quality audio with either TCC or TDC inputs.
\paragraph*{Temporal Controllability}
For temporal controllability, PicoAudio2 markedly outperforms all baselines, including AudioComposer.
Compared with AudioComposer which relies solely on textual descriptions for temporal control, PicoAudio2 leverages a separate timestamp matrix to explicitly encode temporal information, achieving better temporal alignment with the input.
By leveraging the strong language understanding capabilities of LLMs, PicoAudio2 enables controllability not only over timestamps but also over event order, frequency, and other aspects, as illustrated in \Cref{fig:demo}.

\subsection{Ablation Studies}
To verify the necessity of model design and data proposed in PicoAudio2, we further explore the effect of timestamp matrix and real training data.
The timestamp matrix is first excluded from PicoAudio2, denoted as `w/o T'.
This ensembles AudioComposer in that temporal information is included in the caption, so the model relies solely on cross attention with the caption feature to encode temporal information. 
Results in \Cref{tab:main_result} and \ref{tab:audiocaps_result} demonstrate that the exclusion of the timestamp matrix results in inferior temporal controllability, while still maintaining reasonable audio quality, validating the effectiveness of the timestamp matrix.
\begin{table}[h]
 \centering
 \caption{The effect of incorporating real data into training.}
 \label{tab:data_ablation}
 \small
 \begin{tabular}{l|ccc}
 \toprule
 \textbf{Training Data} & \textbf{FD} & \textbf{CLAP} & \textbf{Seg-F$_1$} \\
 \midrule
 \midrule
 Simulation & 41.859 & 0.256 & 0.589 \\
 Simulation + Real  & 39.961  & 0.370& 0.857 \\
 \bottomrule
 \end{tabular}
\end{table}\par
Then, we explore the influence of training data by excluding real data from training.
Results in \Cref{tab:data_ablation} show that training on simulated data alone substantially degrades both audio quality and temporal controllability.
Although event labels are mapped to free-text descriptions to simulate natural language, a significant distribution gap remains between simulated and real data. This discrepancy highlights the critical importance of merging real datasets into training. 

\section{Conclusion}
Mainstream TTA models struggle to achieve high audio quality and precise temporal control on natural language descriptions.
To this end, we propose PicoAudio2 in this work, incorporating new data processing pipelines and framework designs to improve temporal controllable TTA. 
In terms of data, PicoAudio2 designs pipelines for both simulation and real data, obtaining audio-TCC-TDC triplets for training.
In terms of model, PicoAudio2 adopts a new timestamp matrix to represent event captions and their temporal occurrence to provide time-aligned feature for controllable generation.
Objective and subjective evaluation show that PicoAudio2 achieves superior temporal controllability with audio quality comparable to mainstream TTA models. 
Due to challenges in real data annotation, PicoAudio2 is trained only on disjoint subsets of real datasets. It results in limited temporal control on overlapping events, which serves as a direction for our future work.
%

\bibliographystyle{IEEEtran}
\bibliography{refs}

\end{document}